\documentclass[10pt]{article}
\usepackage{graphicx}
\usepackage{amsmath}
\usepackage{amssymb}
\usepackage{caption2}
\begin{document}
\begin{center}
{\large {\bf \sc{Mass spectrum of the hidden-charm hybrid states via the QCD sum rules }}} \\[2mm]
Zhi-Gang  Wang \footnote{E-mail: zgwang@aliyun.com.  }     \\
 Department of Physics, North China Electric Power University, Baoding 071003, P. R. China
\end{center}

\begin{abstract}
In this work,  we  study the mass spectrum of the hidden-charm hybrid states with the $J^{PC}=0^{-+}$, $0^{++}$, $0^{--}$, $1^{++}$, $1^{+-}$, $1^{-+}$, $1^{--}$, $2^{-+}$ and $2^{++}$ via the QCD sum rules in a consistent way. We calculate the vacuum condensates up to dimensions-6  by taking account of both the leading order and next-to-leading order contributions, and  take  the  energy scale formula $\mu=\sqrt{M^2_{X/Y/Z}-(2{\mathbb{M}}_c)^2}$ to choose the suitable energy scales of the QCD spectral densities, it is the first time to explore the energy scale dependence  of the QCD sum rules for the hidden-charm hybrid states.
 \end{abstract}



\section{Introduction}
In the traditional quark model, the hadrons are classified into mesons and baryons, which are bound states of a quark-antiquark pair or three quarks. However, the quark model does not forbid  the possibilities of tetraquark states,  pentaquark states, hybrid states, glue-balls, which are the commonly called  exotic hadrons or $X$, $Y$, $Z$, $P$, $T$ states. The quantum chromodynamics (QCD) allows valence color degrees of freedom, such as hybrid states with gluonic excitations or glue-balls consist of constituent  gluons, which  are a major arena for testing our understanding of the strong interactions beyond perturbative region.

In 2003,  the Belle collaboration observed the first exotic state $X(3872)$ \cite{X3872-2003}, thereafter,  dozens of exotic states have been observed by the  ATLAS, BaBar, Belle,  BESIII, CDF, CMS, D0 and LHCb collaborations \cite{PDG}. The theoretical physicists have proposed many interpretations for the nature of those exotic states, such as the  tetraquark states, pentaquark states, molecular states, hadro-charmonium states, hybrid states, glue-balls, re-scattering effects, etc.
 However, an single theoretical scheme  cannot interpret the entire spectrum of the exotic states satisfactorily due to shortcomings in one way or another.

It has been argued that the $Y(4260)$, $Y(4360)$ and $Y(4140)$ might  be
 hybrid charmonium states $\bar{c}cg$ or their essential components
\cite{Kou:2005gt,Olsen:2017bmm,ZhSL-Y4260,WangZG-Y4140-hybrid,Mahajan-Y4140-hybrid}, however, they have the normal  quantum numbers $J^{PC}=1^{--}$ and $1^{++}$, respectively, just like the traditional charmonium states,  which make the situations  even complex.
In 2021, the LHCb collaboration observed   the $X(4630)$ in the $J/\psi \phi$  mass spectrum with the favored assignment $J^P=1^-$  \cite{LHCb-X4685}. Although its quantum numbers $J^{PC}=1^{-+}$ are exotic, it is not necessary to be a hidden-charm hybrid state, the assignment as a tetraquark state or molecular state with the valence quarks $c\bar{c}s\bar{s}$ is also possible \cite{WZG-NPB-cscs,WangZG-X4630-AHEP}.
We can consult Ref.\cite{Brambilla:2022hhi} for detailed analysis of semi-inclusive decays of the hidden-charm (hidden-bottom) hybrid states to charmonium (bottomonium) states   based on the Born-Oppenheimer effective field theory to diagnose their nature.

At the light sector, there exist  hybrid candidates, such as  the $\pi(1400)$ and  $\pi(1600)$ with the $J^{PC}=1^{-+}$ \cite{PDG}. In 2022, the BESIII collaboration observed  the isoscalar  resonance $\eta(1855)$ with the exotic quantum numbers $J^{PC}=1^{-+}$  in the process $J/\psi
\rightarrow \gamma \eta(1855)\rightarrow \gamma \eta \eta^{\prime }$ \cite{BESIII:2022riz}, it might  be a possible candidate for the  hybrid state
\cite{Chen:2022qpd,Qiu:2022ktc,Shastry:2022mhk}.
Theoretically,  the mass spectrum of the hybrid states have been investigated by
the MIT bag mode \cite{Jaffe:1975fd,Barnes:1982tx,Chanowitz:1982qj},
 the confining linear potential model \cite{Horn:1977rq},
  the flux tube model for QCD \cite{Isgur:1985vy,Close:1994hc,Page:1998gz},
 the QCD sum-rules \cite{Govaerts-NPB-Prop-Gluon,Govaerts-NPB-1985-Hybrid,
Govaerts-NPB-1985-Hybrid-Excit,Govaerts-NPB-Hybrid-Spectrum,
 HuangT-EPJC-1999,Narison-PLB-2000,JinHY-PRD-2003,Narison:2009vj,QiaoCF-JPG-2012,
Steele-JPG-2012,ChenW-JHEP-2013,Chen:2013eha,Huang:2014hya,
Steele-PRD-2018,Barsbay:2022gtu,TangL-PRD-2022,ChenW-PRD-2023,
hybrid-bcg-PLB,ChenHX-PRD-2024,TangL-2024},
 the lattice QCD \cite{HadronSpectrum:2012gic,Latt-JHEP-2016,Ryan:2020iog,Woss:2020ayi,Chen:2022isv}, the Born-Oppenheimer effective field theory \cite{Brambilla:2018pyn,Soto:2023lbh}, etc.

The predictions from different theoretical works differ from each other greatly,
for example, the ground state mass of the $\bar{c}cg$ with the $J^{PC}=1^{-+}$ is
 $3.70\,\rm{GeV}$ \cite{ChenW-JHEP-2013}, $3.93\,\rm{GeV}$ \cite{hybrid-bcg-PLB}, $4.31\,\rm{GeV}$ \cite{Latt-JHEP-2016}, $3.96\,\rm{GeV}$ \cite{Soto:2023lbh},
which makes the relevant problems unresolved  until now, new analysis is necessary and interesting.

The QCD sum rules play an important role in studying the hadron masses, decay constants, form-factors, hadronic coupling constants, etc, and have been applied extensively to study the $X$, $Y$, $Z$, $P$, $T$ states \cite{QCDSR-review-WZG}.
In Ref.\cite{WangHuangTao-3900}, we  explore  the energy scale dependence of the QCD sum rules for the $X$, $Y$ and $Z$ states for the first time, subsequently,   we
suggest an energy scale formula,
\begin{eqnarray}\label{ESF}
\mu&=&\sqrt{M^2_{X/Y/Z}-(2{\mathbb{M}}_Q)^2} \, ,
 \end{eqnarray}
 with the effective heavy quark masses ${\mathbb{M}}_Q$ to obtain the ideal  energy scales  for  the QCD sum rules for the hidden-charm and hidden-bottom tetraquark states \cite{Wang-tetra-formula,Wang-Huang-NPA-2014}, which  can enhance the ground state contributions significantly
 and improve  the convergent behavior  of the operator product expansion significantly. This  is our unique feature.

In our unique  scheme of the QCD sum rules,
we  have performed a systematic  analysis of  the hidden-charm tetraquark states with the  $J^{PC}=0^{++}$,  $0^{-+}$, $0^{--}$, $1^{--}$, $1^{-+}$, $1^{+-}$, $1^{++}$, $2^{++}$ \cite{WZG-NPB-cscs,WZG-tetra-psedo-NPB,WZG-HC-spectrum-PRD,WZG-HC-spectrum-NPB,
WZG-EPJC-P-2P,WZG-NPB-cucd}, hidden-bottom tetraquark states with the  $J^{PC}=0^{++}$, $1^{+-}$, $1^{++}$, $2^{++}$ \cite{WZG-HB-spectrum-EPJC}, hidden-charm molecular states with the  $J^{PC}=0^{++}$, $1^{+-}$, $1^{++}$, $2^{++}$ \cite{WZG-mole-IJMPA}, doubly-charm tetraquark (molecular) states with the  $J^{P}=0^{+}$, $1^{+}$, $2^{+}$ \cite{WZG-tetra-cc-EPJC} (\cite{WZG-XQ-mole-EPJA}), hidden-charm pentaquark (molecular) states with the $J^{P}={\frac{1}{2}}^{-}$, ${\frac{3}{2}}^{-}$, ${\frac{5}{2}}^{-}$ \cite{WZG-penta-cc-IJMPA-2050003}(\cite{XWWang-penta-mole}). In this work, we extend our previous works to study the hidden-charm hybrid states as there exists  the strong fine structure constant $\alpha_s(\mu)=\frac{g_s^2}{4\pi}$ even in the leading order, we should take  account of the energy scale dependence in a consistent way, just like what we have  done in our previous works.

The article is arranged as follows:  we obtain the QCD sum rules for the  hidden-charm hybrid states in section 2; in section 3, we   present the numerical results and discussions; section 4 is reserved for our conclusion.

\section{QCD sum rules for  the  hidden-charm  hybrid  states}
Firstly,  we write down  the two-point correlation functions $\Pi(p)$, $\Pi_{\mu\mu^\prime}(p)$ and $\Pi_{\mu\nu\mu^\prime\nu^\prime}(p)$,
\begin{eqnarray}\label{CF-Pi}
\Pi(p)&=&i\int d^4x e^{ip \cdot x} \langle0|T\Big\{J(x)J^{\dagger}(0)\Big\}|0\rangle \, ,\nonumber\\
\Pi_{\mu\mu^\prime}(p)&=&i\int d^4x e^{ip \cdot x} \langle0|T\Big\{J_{\mu}(x)J_{\mu^\prime}^{\dagger}(0)\Big\}|0\rangle \, ,\nonumber\\
\Pi_{\mu\nu\mu^\prime\nu^\prime}(p)&=&i\int d^4x e^{ip \cdot x} \langle0|T\Big\{J_{\mu\nu}(x)J_{\mu^\prime\nu^\prime}^{\dagger}(0)\Big\}|0\rangle \, ,
\end{eqnarray}
where  the $T$ denotes the time-ordering operation,  the interpolating currents $J(x)=J^{P}(x)$, $J^S(x)$, $J_\mu(x)=J^{V}_\mu(x)$, $J^A_\mu(x)$,
$J_{\mu\nu}(x)=J_{\mu\nu}^{0}(x)$, $J_{\mu\nu}^{5}(x)$, $J_{\mu\nu}^{\sigma,0}(x)$,
$J_{\mu\nu}^{\sigma,5}(x)$, $J_{\mu\nu}^{2,\sigma,0}(x)$,
$J_{\mu\nu}^{2,\sigma,5}(x)$,
\begin{eqnarray}
J^P(x)&=& \bar{c}_{i}(x)i\gamma_5 \sigma^{\mu\alpha} G^{ij}_{\alpha\mu}(x)  c_j(x)   \, , \nonumber\\
J^S(x)&=& \bar{c}_{i}(x) \sigma^{\mu\alpha} G^{ij}_{\alpha\mu}(x)  c_j(x)   \, ,
\end{eqnarray}
\begin{eqnarray}
J_\mu^V(x)&=& \bar{c}_{i}(x) \gamma^\alpha G^{ij}_{\alpha\mu}(x)  c_j(x)   \, , \nonumber \\
J_\mu^A(x)&=& \bar{c}_{i}(x) \gamma^\alpha \gamma_5 G^{ij}_{\alpha\mu}(x)  c_j(x)   \, ,
\end{eqnarray}
\begin{eqnarray}
J_{\mu\nu}^{0}(x)&=& \bar{c}_{i}(x)  G^{ij}_{\mu\nu}(x) c_j(x)  \, , \nonumber \\
J_{\mu\nu}^{5}(x)&=& \bar{c}_{i}(x) i\gamma_5 G^{ij}_{\mu\nu}(x) c_j(x)  \, ,
\end{eqnarray}
\begin{eqnarray}
J_{\mu\nu}^{\sigma,0}(x)&=& \bar{c}_{i}(x) \left[\sigma_{\mu}{}^\alpha G^{ij}_{\alpha\nu}(x)  -\sigma_{\nu}{}^\alpha  G^{ij}_{\alpha\mu}(x)\right]c_j(x)  \, , \nonumber\\
J_{\mu\nu}^{\sigma,5}(x)&=& \bar{c}_{i}(x) i\gamma_5 \left[\sigma_{\mu}{}^\alpha  G^{ij}_{\alpha\nu}(x)  -\sigma_{\nu}{}^\alpha  G^{ij}_{\alpha\mu}(x)\right]c_j(x)   \, ,
\end{eqnarray}
\begin{eqnarray}
J_{\mu\nu}^{2,\sigma,0}(x)&=& \bar{c}_{i}(x) \left[\sigma_{\mu}{}^\alpha G^{ij}_{\alpha\nu}(x)+\sigma_{\nu}{}^\alpha  G^{ij}_{\alpha\mu}(x)-\frac{1}{2}g_{\mu\nu}\sigma^{\beta\alpha}  G^{ij}_{\alpha\beta}(x)\right]c_j(x)  \, , \nonumber\\
J_{\mu\nu}^{2,\sigma,5}(x)&=& \bar{c}_{i}(x) i\gamma_5 \left[\sigma_{\mu}{}^\alpha  G^{ij}_{\alpha\nu}(x)+\sigma_{\nu}{}^\alpha  G^{ij}_{\alpha\mu}(x)-\frac{1}{2}g_{\mu\nu}\sigma^{\beta\alpha}  G^{ij}_{\alpha\beta}(x)\right]c_j(x)   \, ,
\end{eqnarray}
the subscripts and superscripts $i$ and $j$ of the $c$-quark, $\bar{c}$-quark and gluon  fields are color indexes, the gluon field strength $G^{ij}_{\alpha\beta}=G^a_{\alpha\beta}t^a_{ij}$, $G^a_{\alpha\beta}=\partial_{\alpha}G^a_\beta-\partial_{\alpha}G^a_\beta+g_sf^{abc}G^b_\alpha G^c_\beta$, $t^a=\frac{\lambda^a}{2}$, the $\lambda^a$ is the Gell-Mann matrix.
We modify the hybrid currents in Ref.\cite{Govaerts-NPB-Hybrid-Spectrum} to have definite quantum numbers so as to
avoid using complex projectors to obtain  the hadronic representations.

The  $J^{P}(x)$ and $J^{S}(x)$ couple potentially to the hybrid states with the
$J^{PC}=0^{-+}$ and $0^{++}$, respectively, the superscripts $P$ and $S$ denote pseudoscalar and scalar, respectively.
The  $J_\mu^{V}(x)$ and $J_\mu^{A}(x)$ couple potentially to the hybrid states with the $J^{PC}=1^{-+}$ ($0^{++}$) and $1^{+-}$ ($0^{--}$), respectively, the superscripts $V$ and $A$ denote vector and axialvector, respectively.

The  $J_{\mu\nu}^{0}(x)$ and $J_{\mu\nu}^{5}(x)$ couple potentially  to the hybrid states with the $J^{PC}=1^{+-}$ and $1^{--}$.
The  $J_{\mu\nu}^{\sigma,0}(x)$ and $J_{\mu\nu}^{\sigma,5}(x)$ couple potentially  to the hybrid states with the
$J^{PC}=1^{++}$ and $1^{-+}$. The  $J_{\mu\nu}^{2,\sigma,0}(x)$ and $J_{\mu\nu}^{2,\sigma,5}(x)$ couple potentially  to the hybrid states with the
$J^{PC}=2^{++}$ and $2^{-+}$, respectively. The superscripts $0$, $5$ and $\sigma$ denote that there exists  a Dirac matrix $1$, $\gamma_5$ and $\sigma_{\alpha\beta}$ in the currents, respectively. The superscript $2$ denotes the spin $j=2$.

At the hadron side, we insert  a complete set of intermediate hadronic states with the same quantum numbers  as the  currents $J(x)$, $J_\mu(x)$ and $J_{\mu\nu}(x)$ into the
correlation functions $\Pi(p)$, $\Pi_{\mu\mu^\prime}(p)$ and $\Pi_{\mu\nu\mu^\prime\nu^\prime}(p)$  to obtain the hadronic representation, and isolate the ground state (in other words, pole) contributions \cite{SVZ79,Reinders85},
\begin{eqnarray}
\Pi(p)&=&\frac{\lambda_{P/S}^2}{M_{P/S}^2-p^2}+\cdots \nonumber\\
      &=&\Pi_{P/S}(p^2)\, ,
\end{eqnarray}
\begin{eqnarray}
\Pi_{\mu\mu^\prime}(p)&=&\frac{\lambda_{A/V}^2}{M_{A/V}^2-p^2}\,\tilde{g}_{\mu\mu^\prime}
+\frac{\lambda_{P/S}^2}{M_{P/S}^2-p^2}\,\tilde{p}_{\mu}\tilde{p}_{\mu^\prime}+\cdots \nonumber\\
  &=&\Pi_{A/V}(p^2)\,\tilde{g}_{\mu\mu^\prime}+\Pi_{P/S}(p^2)\,\tilde{p}_{\mu}\tilde{p}_{\mu^\prime}\, ,
\end{eqnarray}
\begin{eqnarray}
\Pi_{\mu\nu\mu^\prime\nu^\prime}^{0}(p)&=&\varepsilon_{\mu\nu\alpha\beta}
\varepsilon_{\mu^\prime\nu^\prime\alpha^\prime\beta^\prime}
\tilde{g}^{\alpha\alpha^\prime}\tilde{p}^{\beta}\tilde{p}^{\beta^\prime}\,\Pi_A(p^2)
+S_{\mu\nu\mu^\prime\nu^\prime}\,\Pi_V(p^2)\, , \nonumber\\
\Pi_{\mu\nu\mu^\prime\nu^\prime}^{5}(p)&=&\varepsilon_{\mu\nu\alpha\beta}
\varepsilon_{\mu^\prime\nu^\prime\alpha^\prime\beta^\prime}
\tilde{g}^{\alpha\alpha^\prime}\tilde{p}^{\beta}\tilde{p}^{\beta^\prime}\,\Pi_V(p^2)
+S_{\mu\nu\mu^\prime\nu^\prime}\,\Pi_A(p^2)\, ,
\end{eqnarray}
\begin{eqnarray}
\Pi_{\mu\nu\mu^\prime\nu^\prime}^{\sigma,0}(p)&=&\varepsilon_{\mu\nu\alpha\beta}
\varepsilon_{\mu^\prime\nu^\prime\alpha^\prime\beta^\prime}
\tilde{g}^{\alpha\alpha^\prime}\tilde{p}^{\beta}\tilde{p}^{\beta^\prime}\,\Pi_A(p^2)
+S_{\mu\nu\mu^\prime\nu^\prime}\,\Pi_V(p^2)\, , \nonumber\\
\Pi_{\mu\nu\mu^\prime\nu^\prime}^{\sigma,5}(p)&=&\varepsilon_{\mu\nu\alpha\beta}
\varepsilon_{\mu^\prime\nu^\prime\alpha^\prime\beta^\prime}
\tilde{g}^{\alpha\alpha^\prime}\tilde{p}^{\beta}\tilde{p}^{\beta^\prime}\,\Pi_V(p^2)
+S_{\mu\nu\mu^\prime\nu^\prime}\,\Pi_A(p^2)\, ,
\end{eqnarray}
\begin{eqnarray}
\Pi_{\mu\nu\mu^\prime\nu^\prime}^{2,\sigma,0/5}(p)&=&\frac{\lambda_T^2}{M_T^2-p^2}
\left(\frac{\tilde{g}_{\mu\mu^\prime}\tilde{g}_{\nu\nu^\prime}+
\tilde{g}_{\nu\mu^\prime}\tilde{g}_{\mu\nu^\prime}}{2}
-\frac{\tilde{g}_{\mu\nu}\tilde{g}_{\mu^\prime\nu^\prime}}{3} \right)+\cdots\, ,\nonumber\\
&=&\Pi_{T}(p^2)
\left(\frac{\tilde{g}_{\mu\mu^\prime}\tilde{g}_{\nu\nu^\prime}+
\tilde{g}_{\nu\mu^\prime}\tilde{g}_{\mu\nu^\prime}}{2}
-\frac{\tilde{g}_{\mu\nu}\tilde{g}_{\mu^\prime\nu^\prime}}{3} \right)+\cdots\, ,
\end{eqnarray}
\begin{eqnarray}
S_{\mu\nu\mu^\prime\nu^\prime}&=&\tilde{g}_{\mu\mu^\prime}\tilde{p}_{\nu}\tilde{p}_{\nu^\prime} -\tilde{g}_{\nu\mu^\prime}\tilde{p}_{\mu}\tilde{p}_{\nu^\prime}
-\tilde{g}_{\mu\nu^\prime}\tilde{p}_{\nu}\tilde{p}_{\mu^\prime}
+\tilde{g}_{\nu\nu^\prime}\tilde{p}_{\mu}\tilde{p}_{\mu^\prime}\, ,
\end{eqnarray}
where we have taken the following  definitions for the pole residues and polarization vectors, \begin{eqnarray}
\langle 0|J^{P}(0)|H_{0^{-+}}(p)\rangle &=&\lambda_{H} \, , \nonumber \\
\langle 0|J^{S}(0)|H_{0^{++}}(p)\rangle &=&\lambda_{H} \, ,
\end{eqnarray}
\begin{eqnarray}
\langle 0|J_\mu^V(0)|H_{1^{-+}}(p)\rangle &=&\lambda_{H}\varepsilon_\mu\, , \nonumber \\
\langle 0|J_\mu^A(0)|H_{1^{+-}}(p)\rangle &=&\lambda_{H}\varepsilon_\mu\, ,
\end{eqnarray}
\begin{eqnarray}
\langle 0|J_\mu^V(0)|H_{0^{++}}(p)\rangle &=&\lambda_{H}\tilde{p}_\mu\, , \nonumber \\
\langle 0|J_\mu^A(0)|H_{0^{--}}(p)\rangle &=&\lambda_{H}\tilde{p}_\mu\, ,
\end{eqnarray}
\begin{eqnarray}
\langle 0|J_{\mu\nu}^{0}(0)|H_{1^{+-}}(p)\rangle &=&\lambda_{H}\varepsilon_{\mu\nu\alpha\beta}\,\tilde{p}^\alpha\varepsilon^\beta\, ,\nonumber \\
\langle 0|J_{\mu\nu}^{0}(0)|H_{1^{--}}(p)\rangle &=&\lambda_{H}\left(\tilde{p}_\mu\varepsilon_\nu-\tilde{p}_\nu\varepsilon_\mu\right)\, ,
\end{eqnarray}
\begin{eqnarray}
\langle 0|J_{\mu\nu}^{5}(0)|H_{1^{+-}}(p)\rangle &=&\lambda_{H}\left(\tilde{p}_\mu\varepsilon_\nu-\tilde{p}_\nu\varepsilon_\mu\right)\, ,\nonumber \\
\langle 0|J_{\mu\nu}^{5}(0)|H_{1^{--}}(p)\rangle &=&\lambda_{H}\varepsilon_{\mu\nu\alpha\beta}\,\tilde{p}^\alpha\varepsilon^\beta\, ,
\end{eqnarray}
\begin{eqnarray}
\langle 0|J_{\mu\nu}^{\sigma,0}(0)|H_{1^{++}}(p)\rangle &=&\lambda_{H}\varepsilon_{\mu\nu\alpha\beta}\,\tilde{p}^\alpha\varepsilon^\beta\, , \nonumber \\
\langle 0|J_{\mu\nu}^{\sigma,0}(0)|H_{1^{-+}}(p)\rangle &=&\lambda_{H}\left(\tilde{p}_\mu\varepsilon_\nu-\tilde{p}_\nu\varepsilon_\mu\right)\, ,
\end{eqnarray}
\begin{eqnarray}
\langle 0|J_{\mu\nu}^{\sigma,5}(0)|H_{1^{++}}(p)\rangle &=&\lambda_{H}\left(\tilde{p}_\mu\varepsilon_\nu-\tilde{p}_\nu\varepsilon_\mu\right)\, , \nonumber \\
\langle 0|J_{\mu\nu}^{\sigma,5}(0)|H_{1^{-+}}(p)\rangle &=&\lambda_{H}\varepsilon_{\mu\nu\alpha\beta}\,\tilde{p}^\alpha\varepsilon^\beta\, ,
\end{eqnarray}
\begin{eqnarray}
\langle 0|J_{\mu\nu}^{2,\sigma,5}(0)|H_{2^{-+}}(p)\rangle &=&\lambda_{H}\varepsilon_{\mu\nu}\, , \nonumber \\
\langle 0|J_{\mu\nu}^{2,\sigma,0}(0)|H_{2^{++}}(p)\rangle &=&\lambda_{H}\varepsilon_{\mu\nu}\, ,
\end{eqnarray}
$\tilde{g}_{\mu\mu^\prime}=-g_{\mu\mu^\prime}+\tilde{p}_{\mu}\tilde{p}_{\mu^\prime}$, $\tilde{p}_{\mu}\tilde{p}_{\mu^\prime}=\frac{p_{\mu}p_{\mu^\prime}}{p^2}$, and the symbols $H=P$, $S$, $V$, $A$ and $T$ denote the pseudoscalar, scalar, vector, axialvector and tensor hybrid states, respectively.   We add  the subscripts  $0^{++}$, $0^{-+}$, $0^{--}$, $1^{++}$, $1^{-+}$, $1^{+-}$, $1^{--}$, $2^{-+}$ and $2^{++}$ to denote the corresponding quantum numbers $J^{PC}$ of the hidden-charm  hybrid states.

At the QCD side,  we contract the quark and gluon fields in the correlation functions $\Pi(p)$, $\Pi_{\mu\mu^\prime}(p)$ and $\Pi_{\mu\nu\mu^\prime\nu^\prime}(p)$ with the Wick theorem, obtain the results, for example,
\begin{eqnarray}\label{CF-VS}
\Pi^V_{\mu\mu^\prime}(p)&=&-ig_s^2 t^a_{ij}t^b_{j^\prime i^\prime} \int d^4x e^{ip \cdot x}  {\rm Tr}\left[ \gamma^\alpha S_c^{jj^\prime}(x)\gamma^\beta S_c^{i^{\prime}i}(-x)\right] S^{ab}_{\alpha\mu\beta\mu^\prime}(x) \, ,
\end{eqnarray}
where the $S^{ab}_{\mu\nu\alpha\beta}(x)$ and $S_c^{ij}(x)$ are the full gluon   and $c$ quark propagators, respectively,
\begin{eqnarray}
S^{ab}_{\mu\nu\alpha\beta}(x)&=& \frac{\delta_{ab}}{2\pi^2 x^6}
\left[g_{\mu\alpha}\left(x^2g_{\nu\beta}-4x_{\nu}x_{\beta}\right)
+g_{\nu\beta}\left(x^2g_{\mu\alpha}-4x_{\mu}x_{\alpha} \right) -g_{\mu\beta}\left(x^2g_{\nu\alpha}-4x_{\nu}x_{\alpha}\right)\right.\nonumber\\
&& \left. -g_{\nu\alpha}\left(x^2g_{\mu\beta}-4x_{\mu}x_{\beta}\right)\right]
-\frac{g_sf^{abc}}{4\pi^2x^4} \left[G^c_{\mu\alpha}\left(x^2g_{\nu\beta}-2x_{\nu}x_{\beta}\right) +G^c_{\nu\beta}\left(x^2g_{\mu\alpha}-2x_{\mu}x_{\alpha}\right)\right. \nonumber\\
&&\left.-G^c_{\mu\beta}\left(x^2g_{\nu\alpha}-2x_{\nu}x_{\alpha}\right)
-G^c_{\nu\alpha}\left(x^2g_{\mu\beta}-2x_{\mu}x_{\beta}\right)   \right]
-\frac{g_sf^{abc}}{8\pi^2x^4} \left[g_{\mu\alpha}G^c_{\lambda\beta}\left(x^2g_{\nu}{}^{\lambda}-2x_{\nu}x^{\lambda}\right) \right.\nonumber\\
&&\left.+g_{\nu\beta}G^c_{\lambda\alpha}\left(x^2g_{\mu}{}^{\lambda}-2x_{\mu}x^{\lambda}\right)  -g_{\mu\beta}G^c_{\lambda\alpha}\left(x^2g_{\nu}{}^{\lambda}-2x_{\nu}x^{\lambda}\right) -g_{\nu\alpha}G^c_{\lambda\beta}\left(x^2g_{\mu}{}^{\lambda}-2x_{\mu}x^{\lambda}\right) \right]\nonumber\\
&&+\cdots\, ,
 \end{eqnarray}
\begin{eqnarray}\label{c-quark-prog}
S_c^{ij}(x)&=&\frac{i}{(2\pi)^4}\int d^4k e^{-ik \cdot x} \left\{
\frac{\delta_{ij}}{\!\not\!{k}-m_c}
-\frac{g_sG^n_{\alpha\beta}t^n_{ij}}{4}\frac{\sigma^{\alpha\beta}(\!\not\!{k}+m_c)
+(\!\not\!{k}+m_c)
\sigma^{\alpha\beta}}{(k^2-m_c^2)^2}\right.\nonumber\\
&&\left. +\frac{g_s D_\alpha G^n_{\beta\lambda}t^n_{ij}(f^{\lambda\beta\alpha}+f^{\lambda\alpha\beta}) }{3(k^2-m_c^2)^4}-\frac{g_s^2 (t^at^b)_{ij} G^a_{\alpha\beta}G^b_{\mu\nu}(f^{\alpha\beta\mu\nu}
+f^{\alpha\mu\beta\nu}+f^{\alpha\mu\nu\beta}) }{4(k^2-m_c^2)^5}\right. \nonumber\\
&&\left.+\frac{i\left(D_{\alpha}D_{\beta}+D_{\beta}D_{\alpha} \right) g_sG^n_{\rho\mu}t^n_{ij}(f^{\mu\rho\alpha\beta}+f^{\mu\alpha\rho\beta}
+f^{\mu\alpha\beta\rho}) }{8(k^2-m_c^2)^5}\right. \nonumber\\
&&\left.+\frac{\langle g_s^3GGG\rangle \left(\!\not\!{k}+m_c\right)\left[\!\not\!{k}\left( k^2-3m_c^2\right) +2m_c\left(2k^2-m_c^2 \right)\right] \left(\!\not\!{k}+m_c\right)}{48(k^2-m_c^2)^6} +\cdots\right\}\, ,
\end{eqnarray}
\begin{eqnarray}\label{f-f-f}
f^{\lambda\alpha\beta}&=&(\!\not\!{k}+m_c)\gamma^\lambda(\!\not\!{k}+m_c)
\gamma^\alpha(\!\not\!{k}+m_c)\gamma^\beta(\!\not\!{k}+m_c)\, ,\nonumber\\
f^{\alpha\beta\mu\nu}&=&(\!\not\!{k}+m_c)\gamma^\alpha(\!\not\!{k}+m_c)
\gamma^\beta(\!\not\!{k}+m_c)\gamma^\mu(\!\not\!{k}+m_c)\gamma^\nu(\!\not\!{k}+m_c)\, ,
\end{eqnarray}
and   $D_\alpha=\partial_\alpha-ig_sG^a_\alpha t^a$, $\langle g_s^3 GGG\rangle=\langle g_s^3 f^{abc}G^a_{\mu\nu}G_{b}^{\nu\alpha}G^c_{\alpha}{}^\mu\rangle$ \cite{Govaerts-NPB-Prop-Gluon,Reinders85}, we add the superscript $V$ in the correlation function to denote the current $J_\mu^V(x)$.
Then we compute the integrals in the coordinate space  and momentum space sequentially in the $D$-dimension, and obtain the QCD spectral densities $\rho_{QCD}(s)$ through dispersion relation. For a detailed example, see Ref.\cite{QCDSR-review-WZG}.
 We consider  the vacuum condensates up to dimension $6$, and compute the vacuum condensates $\langle\frac{\alpha_{s}GG}{\pi}\rangle$, $\langle g_{s}^3 GGG\rangle$ and $\langle\bar{q}q\rangle^2$ with $q=u$, $d$ or $s$.
In calculations, we have used the following formulas,
\begin{eqnarray}
\langle g_s^2 D_\alpha G^a_{\mu\nu} D_\beta G^a_{\rho\sigma}\rangle &=&\frac{g_s^4 \langle jj\rangle}{36}\left(g_{\alpha\nu}g_{\beta\sigma}g_{\mu\rho}-g_{\alpha\mu}g_{\beta\sigma}g_{\nu\rho}
-g_{\alpha\nu}g_{\beta\rho}g_{\mu\sigma}+g_{\alpha\mu}g_{\beta\rho}g_{\nu\sigma} \right)\, ,
\end{eqnarray}
\begin{eqnarray}
\langle g_s^2D_{\alpha}D_{\beta} G^{a}_{\mu\nu}G^a_{\rho\sigma}\rangle &=& \left[ \frac{5\langle g_s^3GGG\rangle}{72}-\frac{g_s^4\langle j j\rangle }{36} \right]g_{\alpha\beta}\left(g_{\mu\rho}g_{\nu\sigma}-g_{\mu\sigma}g_{\nu\rho} \right) \nonumber\\
&&-\left[ \frac{\langle g_s^3GGG\rangle}{144}+\frac{g_s^4\langle j j\rangle}{72} \right]\left[g_{\beta\mu}\left(g_{\alpha\rho}g_{\nu\sigma}
-g_{\alpha\sigma}g_{\nu\rho} \right)-g_{\beta\nu}\left(g_{\alpha\rho}g_{\mu\sigma}
-g_{\alpha\sigma}g_{\mu\rho} \right) \right] \nonumber\\
&&-\left[ \frac{7\langle g_s^3GGG\rangle}{144}+\frac{g_s^4\langle j j\rangle}{72} \right]\left[g_{\alpha\mu}\left(g_{\beta\rho}g_{\nu\sigma}
-g_{\beta\sigma}g_{\nu\rho} \right)-g_{\alpha\nu}\left(g_{\beta\rho}g_{\mu\sigma}
-g_{\beta\sigma}g_{\mu\rho} \right) \right] \, ,\nonumber\\
\end{eqnarray}
\begin{eqnarray}
\langle jj\rangle &=&\langle \bar{\psi}\gamma_\mu t^a \psi\bar{\psi}\gamma^\mu t^a \psi\rangle=-\frac{4}{9}\left[ \langle \bar{u}u\rangle^2+\langle \bar{d}d\rangle^2 +\langle \bar{s}s\rangle^2\right]\, ,
\end{eqnarray}
where $\psi=u$, $d$ and $s$.
The QCD spectral densities $\rho_{QCD}(s)$ have the terms
$\frac{\alpha_s}{\pi}$, $\langle\frac{\alpha_s}{\pi} GG\rangle$, $\langle g_s^3 GGG\rangle$ and $g_s^4\langle jj\rangle$ of the leading-order (LO), and the terms $\frac{\alpha_s}{\pi}\langle\frac{\alpha_s}{\pi} GG\rangle$, $\frac{\alpha_s}{\pi}\langle g_s^3 GGG\rangle$ and $\frac{\alpha_s}{\pi}g_s^4\langle jj\rangle$ of the next-to-leading order (NLO), they are all originated from directly calculating the integrals in Eqs.\eqref{CF-VS}-\eqref{f-f-f}.
Compared with the previous works \cite{QiaoCF-JPG-2012,Steele-JPG-2012,ChenW-JHEP-2013,Steele-PRD-2018,hybrid-bcg-PLB},
we perform the operator product expansion in a more comprehensive way by taking account of both the LO and NLO contributions, as the derivatives  $D_\alpha G^a_{\mu\nu}$ and $D_\alpha D_\beta G^a_{\mu\nu}$ lead to  both the LO and NLO contributions. In Ref.\cite{hybrid-bcg-PLB}, although the higher dimensional condensates $\langle\frac{\alpha_s}{\pi} GG\rangle^2$, $\langle\frac{\alpha_s}{\pi} GG\rangle\langle g_s^3 GGG\rangle$ and $\langle g_s^3 GGG\rangle^2$ are taken into account, they are far from complete, many other vacuum condensates  of dimension-8 are neglected \cite{Reinders85}, furthermore, the important vacuum condensate $\langle jj\rangle$ of dimension-6 is also neglected.

We match  the hadronic representation  with the QCD  representation  for the components $\Pi_{i}(p^2)$ with $i=P$, $S$, $V$, $A$ and $T$  below the continuum thresholds   $s_0$ and accomplish  the Borel transformation  with respect to
the variable $P^2=-p^2$ to obtain   the  QCD sum rules:
\begin{eqnarray}\label{QCDSR}
\lambda^2_{H}\, \exp\left(-\frac{M^2_{H}}{T^2}\right)= \int_{4m_c^2}^{s_0} ds\, \rho_{QCD}(s) \, \exp\left(-\frac{s}{T^2}\right) \, ,
\end{eqnarray}
where the $T^2$ is the Borel parameter.

At last, we differentiate the  QCD sum rules in Eq.\eqref{QCDSR} with respect to  the variable $\tau=\frac{1}{T^2}$,  and obtain the QCD sum rules for  the masses of the hidden-charm hybrid states $H$,
 \begin{eqnarray}\label{Mass-SR}
 M^2_{H}&=& -\frac{\int_{4m_c^2}^{s_0} ds\frac{d}{d \tau}\rho_{QCD}(s)\exp\left(-\tau s \right)}{\int_{4m_c^2}^{s_0} ds \rho_{QCD}(s)\exp\left(-\tau s\right)}\, .
\end{eqnarray}

\section{Numerical results and discussions}
We write down the energy-scale dependence of  the input parameters,
\begin{eqnarray}
\langle\bar{q}q \rangle(\mu)&=&\langle\bar{q}q \rangle({\rm 1GeV})\left[\frac{\alpha_{s}({\rm 1GeV})}{\alpha_{s}(\mu)}\right]^{\frac{12}{33-2n_f}}\, , \nonumber\\
 m_c(\mu)&=&m_c(m_c)\left[\frac{\alpha_{s}(\mu)}{\alpha_{s}(m_c)}\right]^{\frac{12}{33-2n_f}} \, ,\nonumber\\
\alpha_s(\mu)&=&\frac{1}{b_0t}\left[1-\frac{b_1}{b_0^2}\frac{\log t}{t} +\frac{b_1^2(\log^2{t}-\log{t}-1)+b_0b_2}{b_0^4t^2}\right]\, ,
\end{eqnarray}
 where the quarks $q=u$, $d$ and $s$,  $t=\log \frac{\mu^2}{\Lambda_{QCD}^2}$, $b_0=\frac{33-2n_f}{12\pi}$, $b_1=\frac{153-19n_f}{24\pi^2}$, $b_2=\frac{2857-\frac{5033}{9}n_f+\frac{325}{27}n_f^2}{128\pi^3}$,  $\Lambda_{QCD}=210\,\rm{MeV}$, $292\,\rm{MeV}$  and  $332\,\rm{MeV}$ for the flavors  $n_f=5$, $4$ and $3$, respectively  \cite{PDG,Narison-mix}. And we choose $n_f=4$ in the present analysis.

 At the initial  points, we take  the standard values  $\langle
\bar{q}q \rangle=-(0.24\pm 0.01\, \rm{GeV})^3$, $\langle
\bar{s} s \rangle=(0.8 \pm 0.1)\langle \bar{q}q \rangle$,
  $\pi\langle \frac{\alpha_s
GG}{\pi}\rangle=(6.40\pm0.30)\,\rm{GeV}^4 $ and $\langle g_s^3GGG\rangle=(8.2\pm1.0){\rm{GeV}^2}\pi\langle \frac{\alpha_s
GG}{\pi}\rangle$    at the particular energy scale  $\mu=1\, \rm{GeV}$ with $q=u$ and $d$
\cite{SVZ79,Reinders85,Colangelo-Review,Narison-2022,Narison-2024}, and take  the
$\overline{MS}$ mass   $m_{c}(m_c)=(1.275\pm0.025)\,\rm{GeV}$  from the Particle Data Group \cite{PDG}. The values of the gluon condensate and three-gluon condensate have been updated from time to time, and change considerably, we choose most recent values \cite{Narison-2022,Narison-2024}. Thereafter, we would like to refer the $c$-quark mass, vacuum condensates and continuum threshold parameters $s_0$ as the input parameters.

In our previous works, we take the  energy scale formula,
\begin{eqnarray}\label{formula}
\mu&=&\sqrt{M^2_{X/Y/Z}-(2{\mathbb{M}}_c)^2}\, ,
 \end{eqnarray}
 to choose the optimal energy scales of the QCD spectral densities for the hidden-charm tetraquark (molecular) states and pentaquark (molecular) states  \cite{WZG-NPB-cscs,WZG-tetra-psedo-NPB,WZG-HC-spectrum-PRD,WZG-HC-spectrum-NPB,
WZG-EPJC-P-2P,WZG-NPB-cucd,WZG-mole-IJMPA,WZG-penta-cc-IJMPA-2050003,XWWang-penta-mole},
 where the effective $c$-quark mass  ${\mathbb{M}}_c=1.82\,\rm{GeV}$ for the diquark type tetraquark and pentaquark states \cite{WZG-NPB-cscs,WZG-tetra-psedo-NPB,WZG-HC-spectrum-PRD,WZG-HC-spectrum-NPB,
WZG-EPJC-P-2P,WZG-NPB-cucd,WZG-penta-cc-IJMPA-2050003}. In this work, we adopt the value ${\mathbb{M}}_c=1.82\,\rm{GeV}$.

As the spectrum of the hidden-charm hybrid
states is rather vague, we have no definite knowledge about the energy gaps between the ground states and first radial excitations. In practical calculations, we  assume the energy gaps are about $0.6\sim 0.7\,\rm{GeV}$, just like in the case of the hidden-charm tetraquark (molecular) states and pentaquark (molecular) states  \cite{WZG-NPB-cscs,WZG-tetra-psedo-NPB,WZG-HC-spectrum-PRD,WZG-HC-spectrum-NPB,
WZG-EPJC-P-2P,WZG-NPB-cucd,WZG-mole-IJMPA,WZG-penta-cc-IJMPA-2050003,
XWWang-penta-mole}, and change  the continuum threshold parameters $s_0$ and Borel parameters $T^2$ to satisfy  the  four   criteria:\\
$\bullet$ Pole  dominance at the hadron  side;\\
$\bullet$ Convergence of the operator product expansion;\\
$\bullet$ Appearance of the  Borel platforms;\\
$\bullet$ Satisfying the energy scale formula,\\
  via trial  and error.

At first, we define the pole contributions (PC),
\begin{eqnarray}
{\rm{PC}}&=&\frac{\int_{4m_{c}^{2}}^{s_{0}}ds\rho_{QCD}\left(s\right)\exp\left(-\frac{s}{T^{2}}\right)} {\int_{4m_{c}^{2}}^{\infty}ds\rho_{QCD}\left(s\right)\exp\left(-\frac{s}{T^{2}}\right)}\, ,
\end{eqnarray}
 and  the contributions of the vacuum condensates $D(n)$ of dimension $n$,
\begin{eqnarray}
D(n)&=&\frac{\int_{4m_{c}^{2}}^{s_{0}}ds\rho_{QCD,n}(s)\exp\left(-\frac{s}{T^{2}}\right)}
{\int_{4m_{c}^{2}}^{s_{0}}ds\rho_{QCD}\left(s\right)\exp\left(-\frac{s}{T^{2}}\right)}\, .
\end{eqnarray}

\begin{figure}
 \centering
 \includegraphics[totalheight=6cm,width=9cm]{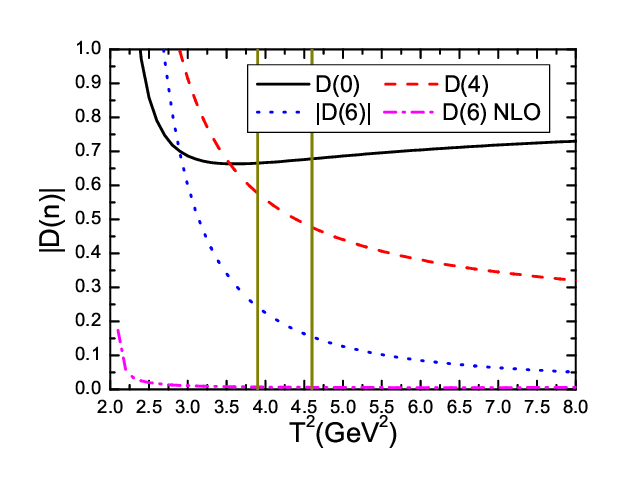}
 \caption{ The contributions of the vacuum condensates for the  hybrid state with the $J^{PC}=1^{-+}$ for the current $J^V_\mu(x)$, where the two vertical lines denote the Borel window.  }\label{fr-J-V-1}
\end{figure}

After numerous  trial and error,  we obtain the Borel windows, continuum threshold parameters, optimal energy scales of the spectral densities and pole contributions, which are  shown explicitly  in Table \ref{BorelP-ss}. At the Borel windows, the  ground state contributions are about $(40-60)\%$, while the central values are slightly larger than $50\%$, the pole dominance criterion  is satisfied, where we have taken the central values of the $c$-quark mass and vacuum condensates. On the other hand, in the Borel windows shown Table \ref{BorelP-ss},  the contributions of the vacuum condensates could be classified into  five relations $D(0)>D(4)\gg |D(6)|$, $D(0)\gg D(4)\gg |D(6)|$, $D(0)\gg |D(4)|\sim |D(6)|$, $D(0)\gg D(4)> |D(6)|$ and $D(0)\sim D(4)\gg |D(6)|$ for the central values of  the input parameters. In all the five cases,  the operator product expansions converge  very well, and we would like to illustrate the first case in Fig.\ref{fr-J-V-1}.

In Fig.\ref{fr-J-V-1}, we plot the contributions of the vacuum condensates for the hidden-charm hybrid state with the $J^{PC}=1^{-+}$ for the current $J^V_\mu(x)$ with variation of the Borel parameter $T^2$ for the central values of the  input  parameters, as an example. From the figure, we can see explicitly that the contributions $D(0)>D(4)\gg |D(6)|$ in the Borel window, on the other hand, the contribution of the $D(6)$ of  the NLO is about $1\%$, as the vacuum condensates of dimension-8 are originated from the operators of higher-order expansion of the full heavy quark propagator, see Eq.\eqref{c-quark-prog}, and companied with the powers $g_s^4$ and $g_s^5$, and their contributions are of the NLO \cite{Reinders85}, and thus they can be neglected safely. All in all, the convergent behaviors of the operator product expansion are very good.

\begin{figure}
 \centering
 \includegraphics[totalheight=6cm,width=9cm]{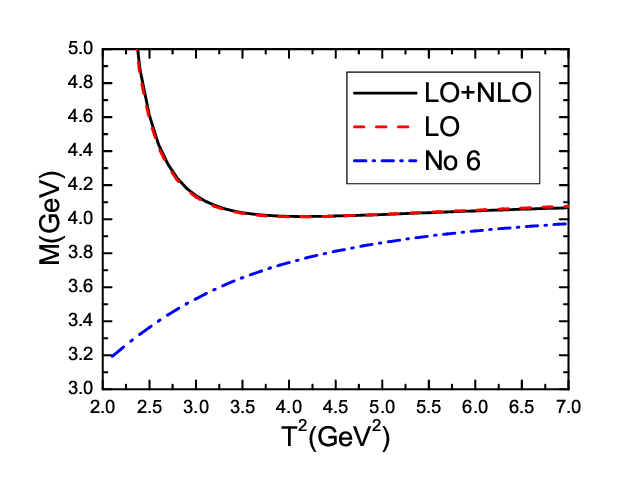}
  \caption{ The mass of the hidden-charm  hybrid state with the $J^{PC}=1^{-+}$ for  the current $J^V_\mu(x)$, where the "No 6" denotes the contributions of the vacuum  condensates of dimension-6 are not included.  }\label{mass-Borel-OPE}
\end{figure}

In Fig.\ref{mass-Borel-OPE}, we plot the mass of the hidden-charm  hybrid state with the $J^{PC}=1^{-+}$ with variation of the Borel parameter $T^2$ for the current $J^V_\mu(x)$ in the cases of different truncations of the operator product expansion for the central values of the input  parameters, as an example. From the figure, we can see explicitly that the NLO contributions can be absorbed into the pole residue safely, and result in almost degenerated mass, while the LO contributions of dimension-6 play an important role and affect the predicted mass significantly beyond the pole residue.

Finally, we take  account of  uncertainties of all the  parameters and obtain  the masses and pole residues of the  hidden-charm hybrid states, which are shown   explicitly in Table \ref{mass-Table-ss}.

In this work,  we calculate the uncertainties $\delta f$  with the
formula,
\begin{eqnarray}
\delta f=\sqrt{\sum_i\left(\frac{\partial f}{\partial
x_i}\right)^2\mid_{x_i=\bar{x}_i} (x_i-\bar{x}_i)^2}\,  ,
\end{eqnarray}
 where the $f$ denotes  the  $M_H$ and $\lambda_H$,  the $x_i$
denotes the  parameters $m_c$,  $\langle \bar{q}q
\rangle$, $\langle \bar{s}s \rangle$,  $\langle \frac{\alpha_s
GG}{\pi}\rangle $, $\langle g_s^3GGG\rangle$,  $s_0$ and  $T^2$. As the partial
 derivatives   $\frac{\partial f}{\partial x_i}$ are difficult to carry
out analytically, we take the  approximation $\left(\frac{\partial
f}{\partial x_i}\right)^2 (x_i-\bar{x}_i)^2\approx
\left[f(\bar{x}_i\pm \delta x_i)-f(\bar{x}_i)\right]^2$ in
numerical calculations.

We obtain the masses of the hidden-charm hybrid states from a fraction, see Eq.\eqref{Mass-SR}, the uncertainties originate from a parameter in the numerator and denominator are canceled  out with each other significantly, the resulting net uncertainties $\delta M_H$ are very small, about $\delta M_H/ M_H\sim (1-3)\%$,  while the uncertainties of the pole residues are larger, about $\delta \lambda_H/ \lambda_H\sim 10\%$, as there no cancelation occurs. The upper  bound $M_H+\delta M_H$ and lower bound $M_H-\delta M_H$ correspond to the continuum threshold parameters $\sqrt{s_0}+\delta\sqrt{s_0}$  and $\sqrt{s_0}-\delta\sqrt{s_0}$, respectively, the relation $\delta M_H\sim \delta\sqrt{s_0}\sim 0.10\,\rm{GeV}$ is roughly satisfied, just like in the case of the hidden-charm tetraquark (molecular) states \cite{WZG-NPB-cscs,WZG-tetra-psedo-NPB,WZG-HC-spectrum-PRD,WZG-HC-spectrum-NPB,
WZG-EPJC-P-2P,WZG-NPB-cucd,WZG-mole-IJMPA}. For example, if we take $\delta\sqrt{s_0}= 0.20\,\rm{GeV}$, then $M_H=4.02\pm 0.12\,\rm{GeV}$ in stead of $4.02\pm 0.08\,\rm{GeV}$ for the current $J_\mu^V(x)$,  the relation $\delta M_H\sim \delta\sqrt{s_0}$ is deviated significantly, the energy gaps $\sqrt{s_0}-M_H $ and $\left[\sqrt{s_0}\pm \delta\sqrt{s_0}\right]-\left[M_H \pm \delta M_H\right]$ would not have consistent values.

From  Tables \ref{BorelP-ss}--\ref{mass-Table-ss}, we  observe clearly    that the  energy scale formula, see Eq.\eqref{formula},  is satisfied very good.

\begin{figure}
 \centering
 \includegraphics[totalheight=5cm,width=6cm]{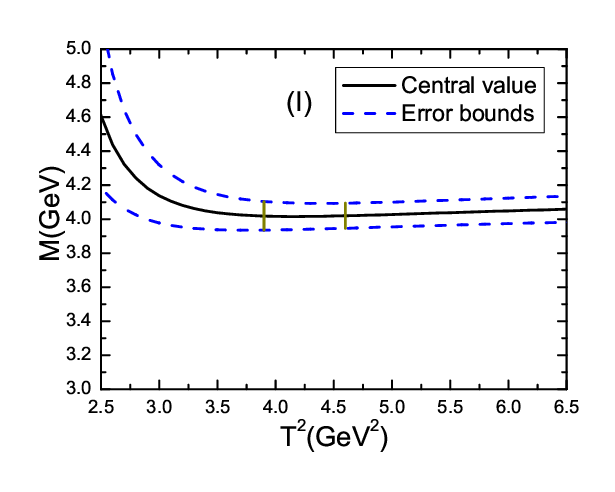}
 \includegraphics[totalheight=5cm,width=6cm]{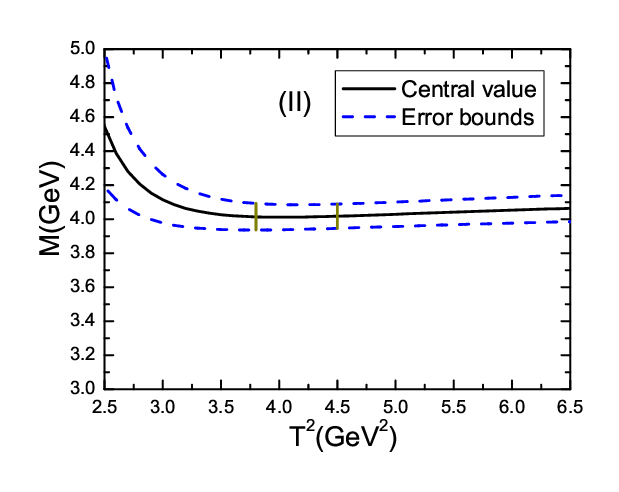}
 \includegraphics[totalheight=5cm,width=6cm]{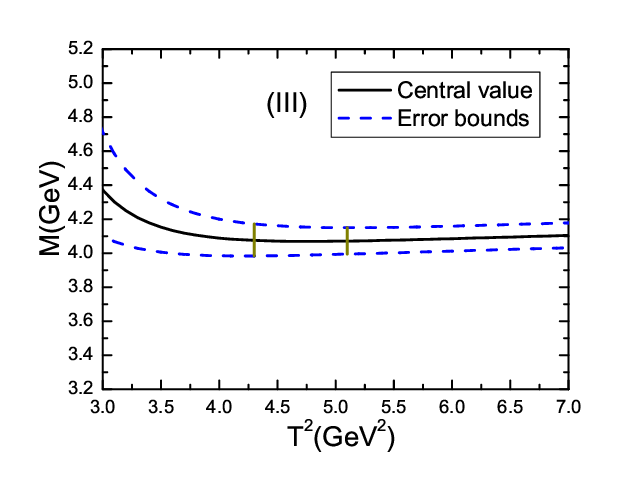}
 \includegraphics[totalheight=5cm,width=6cm]{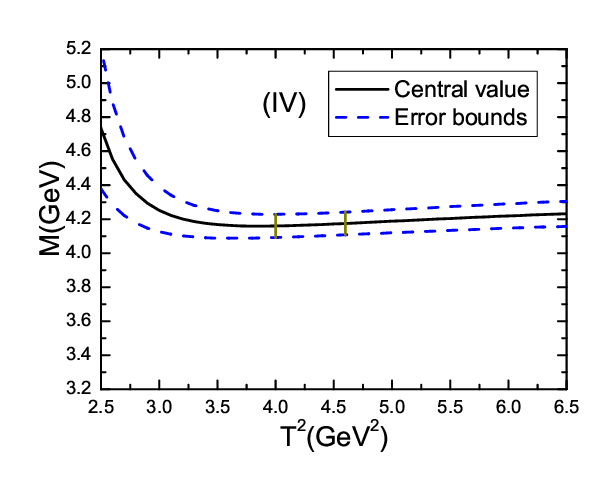}
 \caption{ The masses of the hidden-charm hybrid states, where the (I), (II), (III) and (IV) denote the hybrid states with the $J^{PC}=1^{-+}$, $1^{-+}$, $1^{--}$ and $0^{-+}$ for  the currents $J_\mu^V(x)$, $J_{\mu\nu}^{\sigma,0/5}(x)$, $J^5_{\mu\nu}(x)$ and  $J^P(x)$, respectively, the two vertical lines denote the Borel windows.  }\label{mass-Borel}
\end{figure}

 In  Fig.\ref{mass-Borel}, we plot the masses of the  hidden-charm hybrid states   with  the $J^{PC}=1^{-+}$, $1^{-+}$, $1^{--}$ and $0^{-+}$ for  the currents $J_\mu^V(x)$, $J_{\mu\nu}^{\sigma,0/5}(x)$, $J^5_{\mu\nu}(x)$ and  $J^P(x)$ respectively with variations of the Borel parameters $T^2$, as an example, where the error bounds originate from uncertainties of the input parameters.
 From the figure, we can see explicitly that there really appear elegant  platforms in the Borel windows, the uncertainties come from the Borel parameters are rather small. In fact, we can choose larger Borel parameters at the cost of smaller pole contributions, thus we obtain more flatter platforms and better convergent behavior in the operator product expansion, see Figs.\ref{fr-J-V-1}-\ref{mass-Borel}. Compared with Ref.\cite{ChenW-JHEP-2013}, we choose larger pole contributions, in fact, only the representation in Ref.\cite{ChenW-JHEP-2013} is convenient to compare with.

\begin{figure}
 \centering
 \includegraphics[totalheight=6cm,width=9cm]{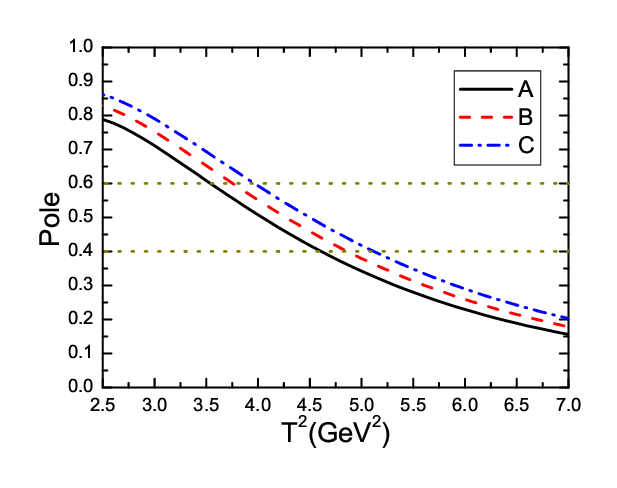}
 \caption{ The pole contribution of the hybrid state with the $J^{PC}=1^{-+}$ for  the current $J^V_\mu(x)$, where the $A$, $B$ and $C$ denote the continuum threshold parameters $\sqrt{s_0}=4.5\,\rm{GeV}$, $4.6\,\rm{GeV}$ and $4.7\,\rm{GeV}$, respectively.  }\label{pole-J-V-1}
\end{figure}

In Fig.\ref{pole-J-V-1}, we plot the pole contribution of the hidden-charm hybrid state with the $J^{PC}=1^{-+}$ for the current $J^V_\mu(x)$ with variation of the Borel parameter $T^2$ for the central values of the $c$-quark mass and vacuum condensates. From the figure, we can see explicitly that the pole contribution decreases monotonically with increase of the Borel parameter, at the value larger than $4.5\,\rm{GeV}^2$, the upper  bound of the Borel parameter, the pole contribution is smaller than $40\%$, although the operator product expansion converges  better, see Fig.\ref{fr-J-V-1}. We prefer larger pole contributions, $(40-60)\%$,  in an uniform way, and expect to obtain robust predictions.

\begin{figure}
 \centering
 \includegraphics[totalheight=4cm,width=6cm]{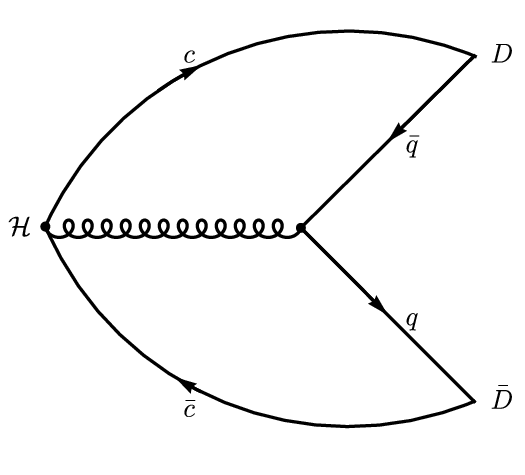}
 \caption{ The Feynman diagram for the  decays of the hidden-charm hybrid states.  }\label{decay-Feynman}
\end{figure}

In Fig.\ref{decay-Feynman}, we draw the Feynman diagram for the  decays of the hidden-charm hybrid states, where the $H$ denotes the hidden-charm hybrid states, the light quarks $q=u$, $d$, $s$, the $D$ denotes the charmed mesons $D$, $D_s$, $D^*$, $D_s^*$, $D_0$, $D_{s0}$, $D_1$, $D_{s1}$, $D_2$ and $D_{s2}$.
 We can take the pole residues $\lambda_H$, see Table \ref{mass-Table-ss}, as  input parameters to explore  the strong decays of those  hidden-charm hybrid states with the three-point QCD sum rules, and obtain ratios among the partial decay widths  to diagnose their nature.  At the present time, the experimental data on the hidden-charm hybrid states are vague, if there really exist a hybrid state with the $J^{PC}=1^{-+}$ at about $4.0\,\rm{GeV}$, see Table \ref{mass-Table-ss}, then the LHCb's new state $X(4630)$ can be tentatively assigned as the first radial excitation according to the energy gap about $0.6\,\rm{GeV}$ \cite{LHCb-X4685}.

 In Ref.\cite{WZG-Landau}, we construct the color singlet-singlet type four-quark current,
 \begin{eqnarray}
J_{\mu\nu}(x)&=&\frac{1}{\sqrt{2}}\Big[\bar{s}(x)\gamma_\mu c(x) \bar{c}(x)\gamma_\nu\gamma_5  s(x)-\bar{s}(x)\gamma_\nu\gamma_5 c(x) \bar{c}(x)\gamma_\mu s(x)\Big]\, ,
 \end{eqnarray}
  to study  the $D_s^*\bar{D}_{s1}-D_{s1}\bar{D}_s^*$ molecular state with the $J^{PC}=1^{-+}$,  and obtain the  prediction,
$M_{X}=4.67\pm0.08\,\rm{GeV}$,  which happens to coincide with the mass of the $X(4630)$ from the LHCb  collaboration, $4626 \pm 16 {}^{+18}_{-110}\,\rm{MeV}$ \cite{LHCb-X4685}, we tentatively assign the $X(4630)$ as the $D_s^*\bar{D}_{s1}-D_{s1}\bar{D}_s^*$ molecular state with the $J^{PC}=1^{-+}$ \cite{WangZG-X4630-AHEP}. In Ref.\cite{WZG-NPB-cscs}, we take the pseudoscalar, scalar,  vector, axialvector and tensor diquarks  as the basic constituents   to construct  the  four-quark currents, and study  the  hidden-charm-hidden-strange tetraquark states with the $J^{PC}=1^{--}$ and $1^{-+}$  comprehensively. According to the predicted mass $4.68\pm0.09\, \rm{GeV}$ of the $[sc]_{S}[\overline{sc}]_{\tilde{V}}+[sc]_{\tilde{V}}[\overline{sc}]_{S}$ state with the   $J^{PC}=1^{-+}$, it is also possible to assign the $X(4630)$ as a tetraquark state.  All in all, the $X(4630)$ might have three important Fock components: hybrid state, molecular state and tetraquark state, we have to study its partial decay widths in details to diagnose its sub-structures, as different sub-structures could lead to quite different partial decay widths.

\begin{table}
\begin{center}
\begin{tabular}{|c|c|c|c|c|c|c|c|c|}\hline\hline
 Currents          &$J^{PC}$ & $T^2 (\rm{GeV}^2)$ & $\sqrt{s_0}(\rm GeV) $      &$\mu(\rm{GeV})$  &pole          \\ \hline

$J^P(x)$        &$0^{-+}$  &$4.0-4.6$  &$4.75\pm0.10$  &$2.0$  &$(41-60)\%$  \\  \hline

$J^S(x)$        &$0^{++}$  &$6.3-7.6$  &$5.75\pm0.10$  &$3.6$  &$(40-61)\%$  \\  \hline

$J^V_{\mu}(x)$  &$0^{++}$  &$7.4-9.0$  &$6.05\pm0.10$  &$3.9$  &$(40-60)\%$  \\  \hline

$J^A_{\mu}(x)$  &$0^{--}$  &$7.8-9.5$  &$6.45\pm0.10$  &$4.5$  &$(40-60)\%$  \\  \hline

$J^V_{\mu}(x)$  &$1^{-+}$  &$3.9-4.6$  &$4.60\pm0.10$  &$1.7$  &$(40-61)\%$  \\  \hline

$J_{\mu\nu}^{\sigma,0/5}(x)$  &$1^{-+}$  &$3.8-4.5$  &$4.60\pm0.10$  &$1.7$  &$(40-61)\%$  \\  \hline

$J_{\mu\nu}^{\sigma,0/5}(x)$  &$1^{++}$  &$6.0-7.3$  &$5.60\pm0.10$  &$3.4$  &$(40-61)\%$  \\  \hline

$J_{\mu\nu}^{0}(x)$  &$1^{+-}$  &$4.8-5.7$  &$5.00\pm0.10$  &$2.4$  &$(40-61)\%$  \\  \hline

$J^A_{\mu}(x)$  &$1^{+-}$  &$5.6-6.7$  &$5.40\pm0.10$  &$3.1$  &$(40-60)\%$  \\  \hline

$J_{\mu\nu}^{5}(x)$  &$1^{+-}$  &$6.6-8.0$  &$5.85\pm0.10$  &$3.7$  &$(40-60)\%$  \\  \hline

$J_{\mu\nu}^{5}(x)$  &$1^{--}$  &$4.3-5.1$  &$4.65\pm0.10$  &$1.8$  &$(41-61)\%$  \\  \hline

$J_{\mu\nu}^{0}(x)$  &$1^{--}$  &$7.3-8.8$  &$6.30\pm0.10$  &$4.3$  &$(40-60)\%$  \\  \hline

$J_{\mu\nu}^{2,\sigma,5}(x)$  &$2^{-+}$  &$4.4-5.2$  &$4.90\pm0.10$  &$2.3$  &$(40-61)\%$  \\  \hline

$J_{\mu\nu}^{2,\sigma,0}(x)$  &$2^{++}$  &$5.6-6.7$  &$5.45\pm0.10$  &$3.2$  &$(40-60)\%$  \\
\hline\hline
\end{tabular}
\end{center}
\caption{ The Borel windows, continuum threshold parameters, energy scales and  pole contributions  for the hidden-charm hybrid states. }\label{BorelP-ss}
\end{table}

\begin{table}
\begin{center}
\begin{tabular}{|c|c|c|c|c|c|c|c|c|}\hline\hline
Currents   &$J^{PC}$  &$M_H (\rm{GeV})$   &$\lambda_H (\rm{GeV}^4) $   \\ \hline

$J^P(x)$        &$0^{-+}$  &$4.17\pm0.08$  &$1.97\pm0.22$  \\ \hline

$J^S(x)$        &$0^{++}$  &$5.10\pm0.06$  &$5.74\pm0.44$  \\ \hline

$J^V_{\mu}(x)$  &$0^{++}$  &$5.37\pm0.06$  &$2.05\pm0.14$  \\ \hline

$J^A_{\mu}(x)$  &$0^{--}$  &$5.79\pm0.06$  &$2.01\pm0.14$  \\ \hline

$J^V_{\mu}(x)$  &$1^{-+}$  &$4.02\pm0.08$  &$(6.18\pm0.64)\times 10^{-1}$  \\ \hline

$J_{\mu\nu}^{\sigma,0/5}(x)$  &$1^{-+}$  &$4.01\pm0.08$  &$(5.80\pm0.62)\times 10^{-1}$  \\ \hline

$J_{\mu\nu}^{\sigma,0/5}(x)$  &$1^{++}$  &$4.96\pm0.06$  &$1.68\pm0.14$  \\ \hline

$J_{\mu\nu}^{0}(x)$  &$1^{+-}$  &$4.36\pm0.09$  &$(4.31\pm0.40)\times 10^{-1}$  \\ \hline

$J^A_{\mu}(x)$  &$1^{+-}$  &$4.76\pm0.07$  &$1.32\pm0.11$  \\ \hline

$J_{\mu\nu}^{5}(x)$  &$1^{+-}$  &$5.21\pm0.07$  &$1.12\pm0.08$  \\ \hline

$J_{\mu\nu}^{5}(x)$  &$1^{--}$  &$4.07\pm0.10$  &$(4.37\pm0.40)\times 10^{-1}$  \\ \hline

$J_{\mu\nu}^{0}(x)$  &$1^{--}$  &$5.61\pm0.07$  &$1.21\pm0.09$  \\ \hline

$J_{\mu\nu}^{2,\sigma,5}(x)$  &$2^{-+}$  &$4.31\pm0.08$  &$1.24\pm0.12$  \\ \hline

$J_{\mu\nu}^{2,\sigma,0}(x)$  &$2^{++}$  &$4.85\pm0.06$  &$2.14\pm0.18$  \\
\hline\hline
\end{tabular}
\end{center}
\caption{ The masses and pole residues of the hidden-charm  hybrid states. }\label{mass-Table-ss}
\end{table}

\section{Conclusion}
In this work,  we extend our previous works on the hidden-charm tetraquark (molecular) states and pentaquark (molecular) states to study the hidden-charm hybrid states with the quantum numbers $J^{PC}=0^{-+}$, $0^{++}$, $0^{--}$, $1^{++}$, $1^{+-}$, $1^{-+}$, $1^{--}$, $2^{-+}$ and $2^{++}$ via the QCD sum rules in an systematic way. We calculate the vacuum condensates up to dimensions six in a consistent way by taking account of both the leading-order and next-to-leading order contributions, and  take  the  energy scale formula $\mu=\sqrt{M^2_{X/Y/Z}-(2{\mathbb{M}}_c)^2}$ to choose the best  energy scales of the QCD spectral densities, it is the first time to explore the energy scale dependence  of the QCD sum rules for the hidden-charm hybrid states. Finally, we obtain the mass spectrum, which can be confronted to experimental  data in the future. While the pole residues can be taken as input parameters to study the two-body strong decays of the hidden-charm hybrid states with the three-point QCD sum rules.

\section*{Acknowledgements}
This  work is supported by National Natural Science Foundation, Grant Number  12175068.

\end{document}